 \documentclass{ws-ijmpa}

\usepackage[super,compress]{cite}
\usepackage{graphicx}

\usepackage{braket}
\usepackage{multirow}
\usepackage{dsfont}
\usepackage{mathtools}
\usepackage{xcolor}
\usepackage[normalem]{ulem}

\newcommand{\bA}{\boldsymbol{A}}
\newcommand{\bt}{\boldsymbol{t}}

\begin{document}
\markboth{Authors' Names}{Instructions for typing manuscripts (paper's title)}

%
\catchline{}{}{}{}{}
%

\title{Three regimes of QCD}

\author{L. Ya. Glozman}

\address{Institute of Physics, University of Graz, A-8010 Graz, Austria}

\maketitle

\begin{history}
\received{Day Month Year}
\revised{Day Month Year}
\end{history}

\begin{abstract}
While the QCD Lagrangian as the whole is only chirally symmetric,
its electric part has larger chiral-spin $SU(2)_{CS}$ and $SU(2N_F)$ symmetries. This allows separation of the electric and magnetic
interactions in a given reference frame. Artificial truncation of the
near-zero modes of the Dirac operator results in the emergence of the 
$SU(2)_{CS}$ and $SU(2N_F)$ symmetries in hadron spectrum. This implies
that  while the confining electric interaction is distributed among
all modes of the Dirac operator, the magnetic interaction is located
at least predominantly in the near-zero modes.  Given this 
observation one could anticipate that above the pseudocritical temperature, where the near-zero modes of the Dirac operator are suppressed, QCD is
$SU(2)_{CS}$ and $SU(2N_F)$ symmetric, which means absence of deconfinement
in this regime. Solution of the $N_F=2$ QCD on the lattice with a chirally 
symmetric Dirac operator reveals that indeed in the interval $Tc - 3Tc$
QCD is approximately $SU(2)_{CS}$ and $SU(2N_F)$ symmetric which implies
that  degrees of freedom are chirally symmetric quarks bound by the
chromoelectric field into color-singlet objects without the chromomagnetic
effects. This regime is referred to as a Stringy Fluid. At larger temperatures
this emergent symmetry smoothly disappears and QCD approaches the Quark-Gluon
Plasma regime with quasifree quarks. The Hadron Gas, the Stringy Fluid and the Quark-Gluon Plasma
differ by symmetries, degrees of freedom and properties.

\end{abstract}



\section{Introduction}	
In this minireview we overview recent results on regimes and symmetries
of QCD at different temperatures. These results are based only upon
a  recently discovered new symmetry of electric interactions in the QCD
lagrangian and its observation in the QCD correlation functions measured
on the lattice at different teperatures. Hence all conclusions are
model independent. It is a standard lore that very soon after the
Big Bang the Universe was a very hot Quark-Gluon Plasma (QGP) that upon
cooling is frosen into hadrons at a pseudocritical temperature
of $T_c \sim  155$ MeV. Here this wisdom is challenged and it is argued
that at temperatures $ T_c - 3T_c$ the hot QCD is not a QGP but a
matter with elementary degrees of freedom  being not liberated
quarks and gluons but rather chirally symmetric quarks bound by the
chromoelectric field into "hadron-like" color-singlet objects.

Asymptotic freedom predicts that at very high temperatures the
strong coupling constant should be sufficiently small so that the
degrees of freedom would be liberated quarks and gluons   \cite{CP}. Such a state of matter was called
the Quark-Gluon Plasma (QGP) \cite{Sh}. For  free quarks chiral symmetry 
is manifest, which is in contrast to the hadronic phase where it is very strongly
dynamically broken. Due to  small coupling constant one can anticipate
that the weak coupling expansion, i.e. the expansion around the point where
quarks and gluons are free particles, should correctly predict properties
of QGP at the sufficiently high temperatures. It was found however very soon
by Linde \cite{L} that at high temperatures even with the small coupling constant there cannot be completely
free deconfined quarks and gluons because of the infrared divergences that 
inevitably appear at some order of perturbation theory. This problem
has not been overcome even today. Inspite of this there was a great
expectation of the community that above some critical temperature $T_c$
QCD is deconfined with restored chiral symmetry so that the degrees of
freedom are free quarks and gluons - the QGP phase. A lot of experimental
and theoretical efforts were (are) invested to discover the Quark-Gluon Plasma.

\begin{figure}
\centering
\includegraphics[angle=0,width=0.45\linewidth]{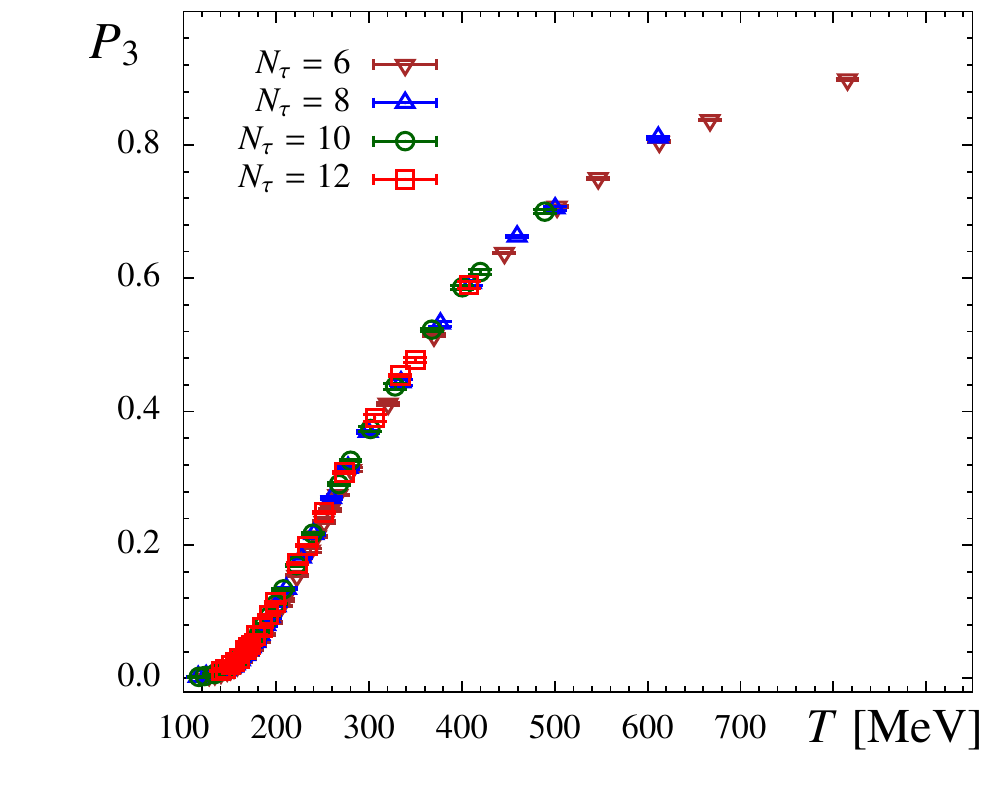}
\caption{ Renormalized Polyakov loop in $N_f = 2 + 1$ QCD at  physical quark masses.
The Fig. is from Ref. \cite{S}.}
\label{St}
\end{figure}

The chiral restoration around $T_c$ was indeed observed on the lattice,
which is signalled by the vanishing of the quark condensate and by the
degeneracy of correlators connected by the chiral transformation \cite{F}.
This restoration happens during a crossover around the pseudocritical
temperature $T_c \sim 155$ MeV.
The expected confinement - deconfinement transition turned out to be
much more difficult to define. For many years such a transition was
believed to be associated with  different behavior of the Polyakov loop
 below and above $T_c$  \cite{P,ML}. The Polyakov loop is
 a true order parameter for the $Z_3$ symmetry of  pure gluodynamics and
 in this case a clear first-order phase transition is observed. If one interprets the Polyakov loop as a free energy of a static charge, then
 a vanishing Polyakov loop below $T_c$ would correspond to a  charge with infinite
 free energy, while a finite Polyakov loop above $T_c$ would represent a 
 charge with a finite free energy. In QCD with dynamical quarks
 there is no $Z_3$ symmetry and the Polyakov loop looses its meaning as
 an order parameter: a static charge interacts with dynamical
 quarks at any temperature and its free energy is not infinite. Indeed in
 QCD with light  quarks one observes nowadays a very smooth increase of
 the Polyakov loop beginning from zero temperature, see Fig ~\ref{St} 
 \cite{S}. 
Obviously no specific temperature of "deconfinement transition"
can be defined.

\begin{figure}
\centering
\includegraphics[angle=0,width=0.45\linewidth]{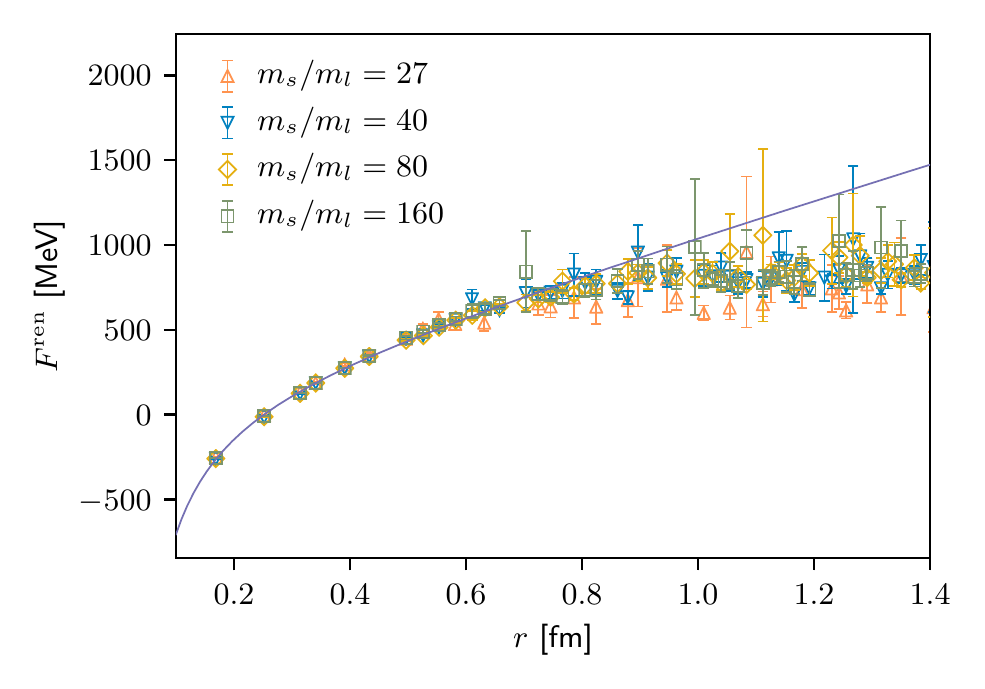}
\includegraphics[angle=0,width=0.45\linewidth]{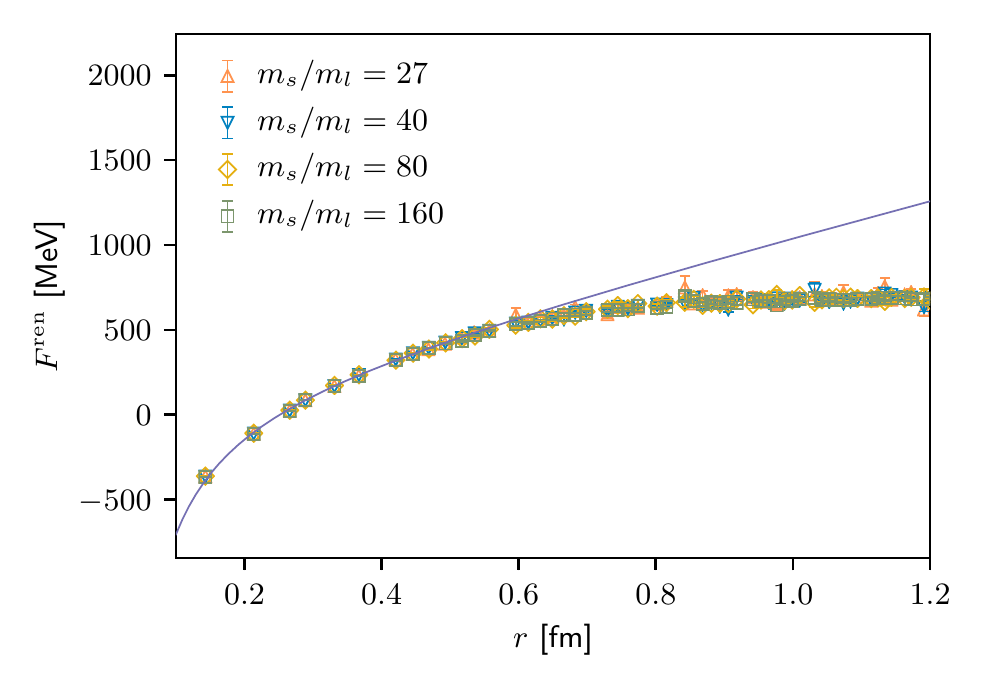}
\caption{ The color-singlet part of the Polyakov loop correlator in 
$N_f = 2 + 1$ QCD at physical and lower than physical quark masses.
The left and right panels are for $T = 141$ MeV and $T = 166$ MeV,
respectively.
The Fig. is from Ref. \cite{K}.}
\label{Karsch}
\end{figure}

  A flattening of a
 potential between the static sources (the Polyakov loop correlator), that is often considered as a sign
 of the Debye screening and deconfinement, is actually induced by the
 "string breaking", i.e. by production of two heavy-light mesons. This can
 be clearly seen from Fig. ~\ref{Karsch}, where such a potential
 is shown for $N_f =2+1$ at physical quark masses \cite{K}. The extracted potential 
 with obvious flattening is exactly the same for temperatures below and above $T_c$. I.e. it cannot be related do "deconfinement".
 
 There is no
 reliable and accepted definition and order parameter for deconfinement
 except for the most straightforward one: confinement is the absence
 of color states in the spectrum. It is also very well established on the
 lattice that only the color-singlet objects survive the gauge averaging
 and consequently only the color-singlet objects can freely
 propagate a macroscopical distance at any temperature.
 
 Tremendous experimental efforts at RHIC and LHC have supplied us
 with  evidences  that indeed at temperatures above $T_c \sim 155$ MeV
 some properties of the QCD matter are different as compared to the hadronic
 matter at small temperatures (perfect fluidity, jet quenching). This, together with the observation
 on the lattice of the pseudocritical temperature  
 was a basis for a 
  declaration of  discovery of the QGP \cite{GL}. Even though that
  the pressure and the energy density immediately above $T_c$ are very far
  from the Stefan-Boltzmann limit that is associated with the free
  quarks and gluons, a general belief was that above $T_c$
  one has the QGP phase with  the liberated quarks and gluons 
  that still undergo  some sufficiently strong interactions - the strongly
  interacting Quark-Gluon Plasma (sQGP). A general justification was 
  that when the energy density is of the order that of matter inside a
  proton (i.e., hadrons overlap) the only possibly is that quarks
  and gluons deconfine in the sense that they freely propagate within
  a macroscopical piece of matter. Such a picture, that was motivated
  by a simple MIT bag model view of hadrons, is implicitly based
  on the assumption that no other possibilities  (than hadrons with broken chiral symmetry or liberated quarks and gluons) exist. Is it however true?
  
  One of the pillars of the QGP  was a very much spread opinion that only free quarks (and no quark bound states) can exist in the chirally symmetric phase and that
   hadron masses, such as of nucleons  etc, are made at least mostly
  from the quark condensate
  of the vacuum. The latter view was a motivation to perform the
  following experiment on
  the lattice \cite{LS,GLS}. The quark condensate of the vacuum is connected to the density
  of the near-zero modes of the Dirac operator via the Bancs-Casher relation
  \cite{BC}. On the lattice one can artificially remove the near-zero modes
  from the quark propagators. Chiral symmetry will be restored by construction.
  But what will happen with hadrons: will they survive  and if yes
  what will be their  masses? It turned out that hadrons (except for pions) 
  do survive this artificial restoration of chiral symmetry and their masses remain large. 
  
  However not only degeneracies in the chiral multiplets were
  observed, but a much larger degeneracy was seen \cite{D1,D2,D3,D4},
  which was very surprising. Given  quantum numbers of degenerate
  hadrons and their  chiral transformation properties
  the unknown symmetry group was reconstructed. It turned out to be the
  $SU(2)_{CS}$ chiral-spin group and its flavor extension $SU(2N_F)$
  \cite{G1,G2}. The $SU(2)_{CS}$ chiral-spin group includes $U(1)_A$
  as a subgroup and its transformations mix the right- and left-handed
  components of quarks. The $SU(2N_F)$ group includes both the chiral
  symmetry $SU(N_F)_R \times SU(N_F)_L \times U(1)_A$ and $SU(2)_{CS}$
  as subgroups. Obviously, these are not symmetries of the Dirac
  equation or of the QCD Lagrangian. However, they are symmetries
  of the Lorentz-invariant fermion charge and of the electric interaction
  in QCD. The magnetic interaction as well as the quark kinetic term
  break them. Consequently these symmetries allow us to distinguish
  between the electric and magnetic interactions in a given reference
  frame. This is a general property of any gauge-invariant theory
  with Dirac particles as a matter field.
  
  The emergence of $SU(2)_{CS}$ and $SU(2N_F)$ symmetries in hadrons
  upon truncation of the near-zero modes of the Dirac operator means
  that while the confining chromoelectric interaction is distributed
  among all modes of the Dirac operator, the chromomagnetic interaction
  contributes only (or at least predominantly) to the near-zero modes.
  Some unknown microscopic dynamics should be responsible for this
  phenomenon. 
  
  A definition of the $SU(2)_{CS}$ and $SU(2N_F)$ symmetries
  as well as results on hadron spectrum upon truncation
  of the near-zero modes at zero temperature will be discussed in detail in sections 2 and 3.
  
  It is known that at high temperatures, above the pseudocritical temperature, the chiral symmetry
  is restored because  the near-zero modes are naturally suppressed
  by temperature. Also the $U(1)_A$ symmetry is at least approximately
  restored \cite{JLQCD1,JLQCD3}. Then, given the experience of
  the emergent symmetries upon the artificial truncation of the
  near-zero modes at zero temperature, one can expect a natural
  emergence of these symmetries above $T_c$ without any truncation
  \cite{G3}. That would imply that there are no free deconfined quarks
  above $T_c$.
  
  This expectation has been tested in lattice simulations of full
  $N_F=2$ QCD with the chirally symmetric Domain Wall Dirac operator.
  Namely, a full set of spatial correlators with $J=0,1$ local meson
  operators has been calculated at temperatures $T_c - 5.5T_c$ 
  \cite{R1,R2}. Approximate degeneracy
  of meson correlators predicted by the $SU(2)_{CS}$ and $SU(2N_F)$ symmetries
  was clearly observed in the temperature interval $T_c - 3 T_c$. 
  The same symmetries have also been seen in temporal correlators that
  are directly connected to observable spectral density \cite{R3}.
  
   The lattice simulations are possible only at zero chemical potential.
  The chemical potential term in the QCD action is manifestly 
  $SU(2)_{CS}$ and $SU(2N_F)$ symmetric \cite{G4} . In other words, the observed
  property of QCD at zero chemical potential should also persist
  at  finite chemical potentials.
  
  These symmetries are incompatible with free deconfined
  quarks and imply that at these temperatures degrees of freedom are
  chirally symmetric quarks bound by the chromoelectric field without
  the chromomagnetic effects into color-singlet compounds. Such a system is reminiscent of a "string",
  that is why this $SU(2)_{CS}$ and $SU(2N_F)$ symmetric regime   of QCD was      named Stringy Fluid \cite{G4}. At larger temperatures $ T > 3 T_c$ these
  symmetries smoothly disappear  and at temperatures $ T \sim 5 T_c$
  only chiral symmetries survive so
  the full QCD correlators approach correlators calculated with free
  noninteracting quarks \cite{R2}. This implies that only at such high
  temperatures the strongly interacting matter can be 
  approximately considered as a Quark-Gluon Plasma with the quasifree quarks  
  and gluons as appropriate degrees of freedom. The Hadron Gas, 
  the Stringy Fluid and the Quark-Gluon Plasma regimes are different  by symmetries, degrees
  of freedom and properties. 
  
  Symmetries of QCD at different temperatures and their
  implications are addressed in detail in sections 4 and
  5.

\section{Chiral-spin symmetry} 
 
 In ref. \cite{G1} the  $SU(2)_{CS}$ chiral-spin transformation 
 was defined as a transformation that rotates in the space of the
 right- and left-handed Weyl spinors

 \begin{equation}
\label{W}
 \left(\begin{array}{c}
R\\
L
\end{array}\right) \rightarrow   \exp \left(i  \frac{\varepsilon^n \sigma^n}{2}\right) \left(\begin{array}{c}
R\\
L
\end{array}\right)\; .
\end{equation}
  So the fundamental irreducible representation of $SU(2)_{CS}$
 is two-dimensional. In terms of the Dirac spinors $\psi$ the same
 transformation can be written via $\gamma$-matrices \cite{G2}

\begin{equation}
\label{V-defsp}
  \psi \rightarrow  \psi^\prime = \exp \left(i  \frac{\varepsilon^n \Sigma^n}{2}\right) \psi = \exp \left(i  \frac{\varepsilon^n \sigma^n}{2}\right) \left(\begin{array}{c}
R\\
L
\end{array}\right)\; ,
\end{equation}

\noindent
where the generators $\Sigma^n$ of the four-dimensional reducible
representation are

\begin{equation}
 \Sigma^n = \{\gamma_0,-i \gamma_5\gamma_0,\gamma_5\},
\label{SIGCS}
\end{equation}
with the $su(2)$ algebra

\begin{equation}
[\Sigma^a,\Sigma^b]=2i\epsilon^{abc}\Sigma^c.
\label{algebra}
\end{equation}
The $U(1)_A$ group is a subgroup of $SU(2)_{CS}$.

In Euclidean space with the $O(4)$ symmetry all four directions are 
equivalent and the $SU(2)_{CS}$ transformations can be generated
by any Euclidean  hermitian $\gamma$-matrix $\gamma_k$, $k=1,2,3,4$ instead
of Minkowskian $\gamma_0$:

\begin{equation}
 \Sigma^n = \{\gamma_k,-i \gamma_5\gamma_k,\gamma_5\},
\label{SIGCS}
\end{equation} 

\begin{equation}
\gamma_i\gamma_j + \gamma_j \gamma_i =
2\delta^{ij}; \qquad \gamma_5 = \gamma_1\gamma_2\gamma_3\gamma_4.
\label{gamma}
\end{equation}
The $su(2)$ algebra 
is satisfied with any $k=1,2,3,4$.

The direct product of the $SU(2)_{CS}$ group with the flavor group
$SU(2)_{CS} \times SU(N_F)$ can be embedded into a $SU(2N_F)$ group.
 This group contains the chiral
symmetry  $SU(N_F)_L \times SU(N_F)_R \times U(1)_A$ as a subgroup.
Its transformations  are given by

\begin{equation}
\psi \rightarrow  \psi^\prime = \exp\left(i \frac{\epsilon^m T^m}{2}\right) \psi, 
\end{equation}

\noindent
where $m=1,2,...,(2N_F)^2-1$ and the set of $(2N_F)^2-1$ generators is

\begin{align}
T^m=\{
(\tau^a \otimes {1}_D),
({1}_F \otimes \Sigma^n),
(\tau^a \otimes \Sigma^n)
\}
\end{align}
with $\tau$  being the flavor generators (with the flavor index $a$) and $n=1,2,3$ is the $SU(2)_{CS}$ index.

The fundamental
vector of $SU(2N_F)$ at $N_F=2$ is

\begin{equation}
\Psi =\begin{pmatrix} u_{\textsc{R}} \\ u_{\textsc{L}}  \\ d_{\textsc{R}}  \\ d_{\textsc{L}} \end{pmatrix}. 
\end{equation}
\noindent

While the $SU(2)_{CS}$ and $SU(2N_F)$ symmetries are not
symmetries of the Dirac Lagrangian, they are symmetries
of the Lorentz-invariant fermion charge
\begin{equation}
Q = \int d^3x \bar \psi(x) \gamma_0 \psi(x) = \int d^3x  
\psi^\dagger(x)  \psi(x).
\label{Q}
\end{equation}

\noindent
The latter important feature allows us to use the 
$SU(2)_{CS}$ and $SU(2N_F)$ symmetries to distinguish the
electric and magnetic interactions in a given reference frame
because  the electric interaction is influenced only by
the charge while the magnetic interaction is dictated by
the spatial current. The latter current is not 
$SU(2)_{CS}$ and $SU(2N_F)$ symmetric. This can be made
explicit as follows.

While in Euclidean space the electric and magnetic fields
cannot be separated because of the $O(4)$ symmetry,
 in Minkowski space in a given reference frame they are
 different fields.
Interaction of  fermions with the gauge field in Minkowski space-time
can be split in a given reference frame into temporal and spatial parts:

\begin{equation}
\overline{\psi}   \gamma^{\mu} D_{\mu} \psi = \overline{\psi}   \gamma^0 D_0  \psi 
  + \overline{\psi}   \gamma^i D_i  \psi ,
\label{cl}
\end{equation}
\noindent
where the covariant derivative $D_{\mu}$  includes
interaction of the matter field $\psi$ with the  gauge field $\bA_\mu$,

\begin{equation}
D_{\mu}\psi =( \partial_\mu - ig \frac{\bt \cdot \bA_\mu}{2})\psi.
\end{equation}
The temporal term contains  interaction of the color-octet
 charge density 

\begin{equation}
\bar \psi (x)  \gamma^0  \frac{\bt}{2} \psi(x) = \psi (x)^\dagger  \frac{\bt}{2} \psi(x)
\label{den}
\end{equation}
with the electric  
part of the gluonic field. 
It is invariant  under 
 $SU(2)_{CS}$  and  $SU(2N_F)$. Note that the $SU(2)_{CS}$ transformations
defined  via the Euclidean
Dirac matrices can be identically applied to Minkowski Dirac spinors without
any modification of the generators.
 
 The spatial part contains the quark kinetic term
and  the interaction with the chromomagnetic field.  It breaks 
 $SU(2)_{CS}$ and $SU(2N_F)$.   We conclude that  interaction
 of the electric and  magnetic components
 of the gauge field with fermions in a given reference frame can be distinguished
 by symmetry. Such a distinction does not exist if the matter
 field is bosonic, because a symmetry of the Klein-Gordon Lagrangian
 and of  charge of the $J=0$ field is the same.

Of course, in order to discuss the electric and magnetic
components of the gauge field
one needs to fix a reference frame. The invariant mass of the hadron
is  the rest frame energy. Consequently, to discuss physics
of hadron mass  it is natural to use the hadron rest frame.
At high temperatures the Lorentz invariance is broken and again
a natural frame  is the medium rest frame.

\section{Emergence of chiral-spin and $SU(2N_F)$
symmetries in hadrons at zero temperature}

The quark condensate of the vacuum is connected with
the density of the near-zero modes of the Euclidean Dirac
operator via the Banks-Casher relation \cite{BC}

\begin{equation}
 <\bar q q> = -\pi \rho(0).
\end{equation} 
 
\begin{figure}
\centering
\includegraphics[angle=0,width=0.45\linewidth]{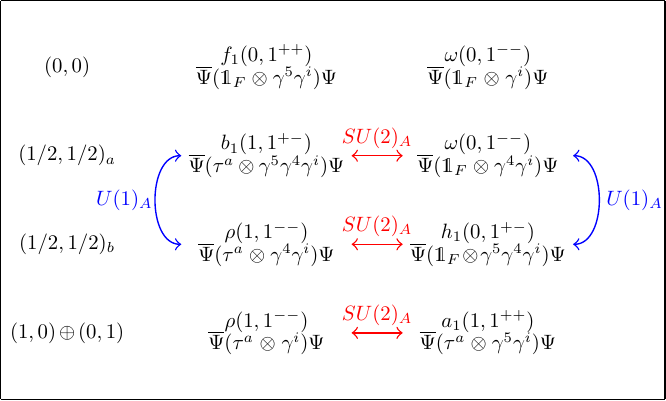}
\includegraphics[angle=0,width=0.45\linewidth]{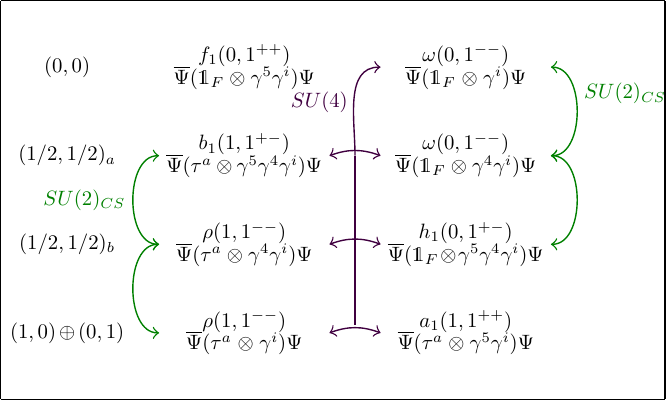}
\caption{Transformations between $J=1$ operators, $i=1,2,3$.
The left columns indicate the $SU(2)_L \times SU(2)_R$ 
representation for every
operator. Red and blue arrows connect operators which transform into 
each other under $SU(2)_L \times SU(2)_R$ and $U(1)_A$, respectively.
Green arrows connect operators that belong to
$SU(2)_{CS}$, $k=4$  triplets. Purple arrow shows the $SU(4)$
15-plet. The $f_1$ operator is is a singlet of $SU(4)$.
The Fig. is from Ref. \cite{G2}.}
\label{F3}
\end{figure}

\noindent
 The hermitian Euclidean Dirac operator, $i \gamma_\mu D_\mu$, 
 has in a finite volume $V$ a discrete spectrum with real eigenvalues $\lambda_n$:

\begin{equation}
i \gamma_\mu D_\mu  \psi_n(x) = \lambda_n \psi_n(x).
\label{ev}
\end{equation}
Consequently, removing by hands $k$ lowest lying modes
of the Dirac operator from the quark propagators,
 
 \begin{equation}
 S =S_{Full}-
  \sum_{i=1}^{k}\,\frac{1}{\lambda_i}\,|\lambda_i\rangle \langle \lambda_i|,
 \end{equation} 
one apriori expects restoration of chiral 
$SU(2)_L \times SU(2)_R$ and possibly of $U(1)_A$
symmetries, if hadrons survive. This should be signalled by degeneracy
of hadrons connected by the $SU(2)_L \times SU(2)_R$ and $U(1)_A$
transformations, see Fig. \ref{F3}.

\begin{figure}
\centering 
\includegraphics[width=0.55\linewidth]{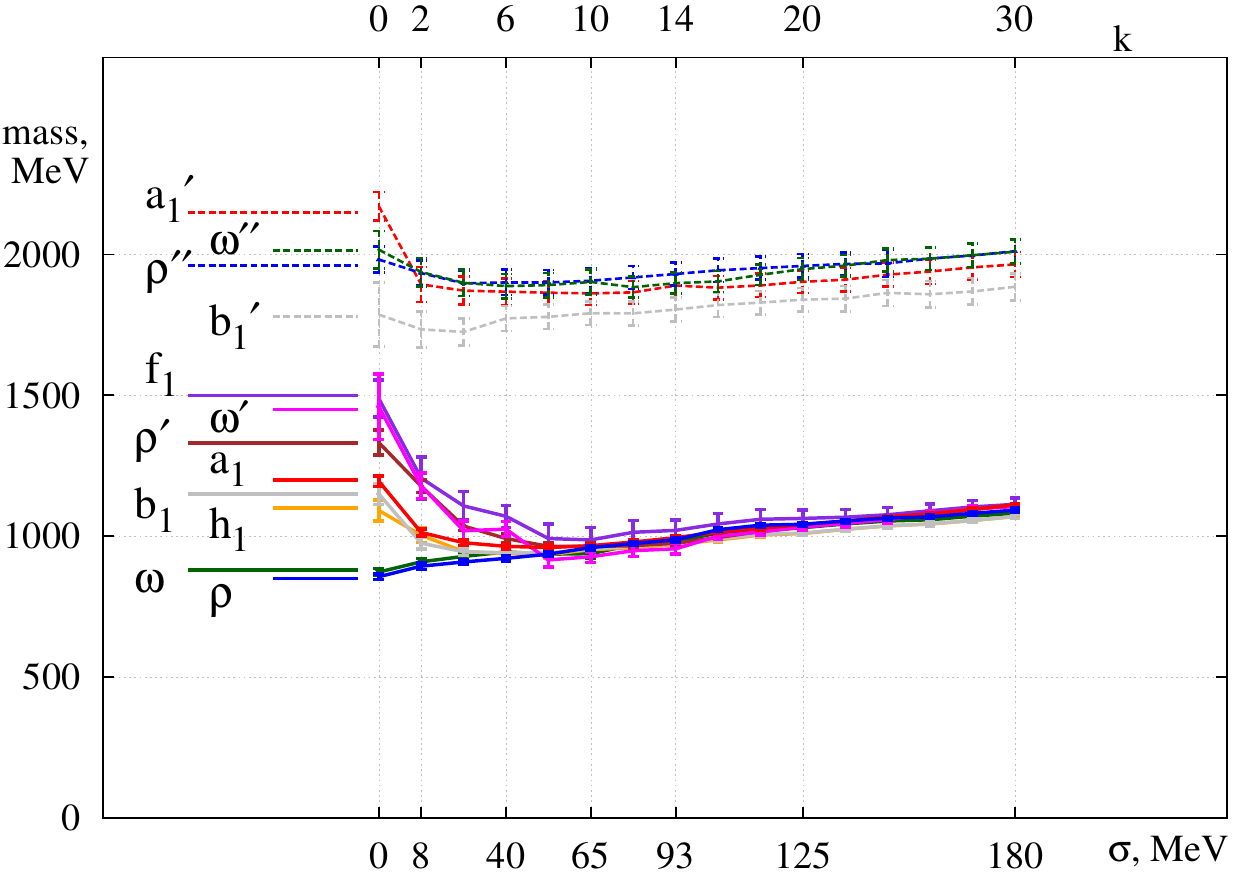}
\caption{$J=1$ meson masses  as a function of the
truncation number $k$ where $k$ represents the amount of removed
lowest modes of the Dirac operator. $\sigma$ shows the energy gap in the Dirac spectrum. The Fig. is from Ref. \cite{D2}.} 
\label{F2}
\end{figure}

Surprisingly a larger degeneracy than the 
$SU(2)_L \times SU(2)_R \times U(1)_A$ symmetry
of the QCD Lagrangian was observed in actual lattice measurements
based on JLQCD overlap gauge configurations
\cite{D1,D2}, see Fig. \ref{F2}.
This large degeneracy, presumably only approximate, represents the $SU(2)_{CS}, ~k=4$ (5) and the $SU(2N_F)$ symmetries since it contains irreducible representations of both groups, see Fig. \ref{F3}.
The $SU(2)_{CS}, ~k=4$
and $SU(2N_F)$ transformations
commute with the $O(3)$ transformations and consequently there  is
no contradiction between the chiral-spin symmetry and conserved
spin of hadrons.

These  results imply, given the symmetry classification of
the QCD Lagrangian, that while the confining chromoelectric
interaction is distributed among
 all modes of the Dirac operator, the
chromomagnetic interaction, which breaks both  symmetries, is located 
at least predominantly in the near-zero
modes. Consequently  artificial removal of the near-zero modes leads to
the emergence of $SU(2)_{CS}$  and $SU(4)$ in hadron spectrum.
Similar
results are also observed in $J=2$ mesons \cite{D3} and baryons \cite{D4}.

The highly degenerate level seen in Fig. \ref{F2} represents a $SU(2N_F)$-
symmetric level of the pure electric confining interaction.\footnote{In
reality the degeneracy of  Fig. \ref{F2} represents a larger
$SU(4) \times SU(4)$ symmetry, see discussion in Ref. \cite{G5}} The hadron
spectra  could be viewed as a  splitting of
the  level of the  QCD string by means of dynamics
contained in the near-zero modes of the Dirac operator, i.e. dynamics
of chiral symmetry breaking that also includes  magnetic effects
in QCD.

\section{Emergence of chiral-spin and $SU(2N_F)$
symmetries at high temperatures}

Above the pseudocritical temperature the chiral symmetry is restored
and consequently the near-zero modes of the Dirac operator
are suppressed by temperature.
Then, given the results on the artificial truncation of the near-zero modes at zero temperature, presented in previous section, one can expect natural emergence of the $SU(2)_{CS}$  and $SU(4)$ symmetries above $T_c$.
This expectation has been verified in the lattice simulations of $N_F=2$
QCD with quarks of the physical mass with the
chirally symmetric Domain Wall Dirac operator (JLQCD ensembles)
at temperatures $T = (220 - 960)$
MeV \cite{R1,R2}. Namely the spatial ($z$ direction) rest-frame correlators of all local isovector $J=0,1$
meson operators  
\begin{equation}
C_\Gamma(n_z) = \sum\limits_{n_x, n_y, n_t}
\braket{\mathcal{O}_\Gamma(n_x,n_y,n_z,n_t)
\mathcal{O}_\Gamma(\mathbf{0},0)^\dagger}
\label{eq:momentumprojection}
\end{equation} 
have been calculated as functions of the dimensionless
variable  
\begin{align}
z\,T \; = \; (n_z a)/(N_t a) \; = \; n_z/N_t \; ,
\label{z_dimless}
\end{align}
where~$z$ is the physical distance in the correlators, $T$ the temperature,
$a$ the lattice constant, $n_z$ the distance in lattice units and $N_t$ the
temporal lattice extent. 

\begin{figure}
  \centering
  \includegraphics[scale=0.4]{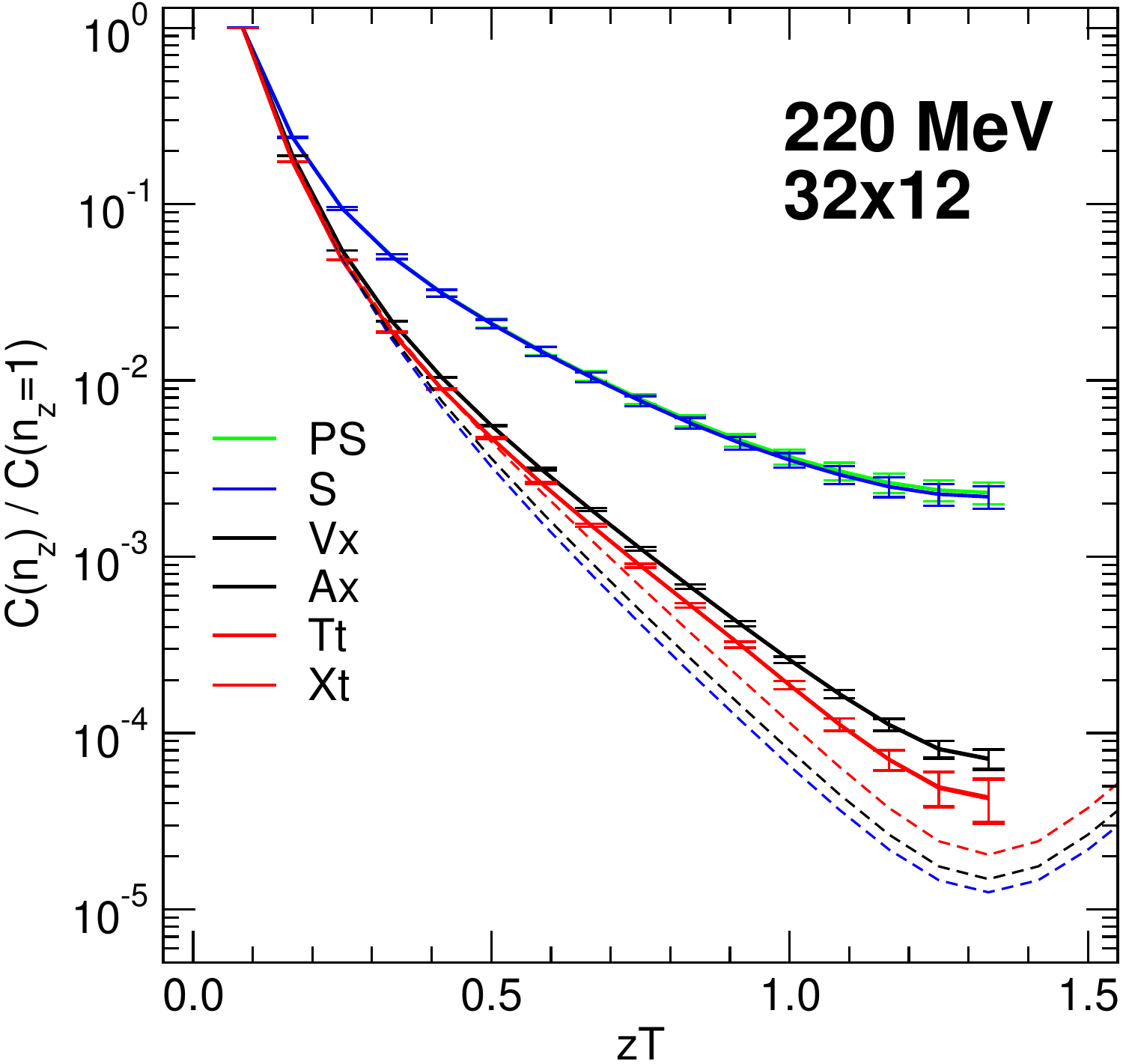} 
  \includegraphics[scale=0.4]{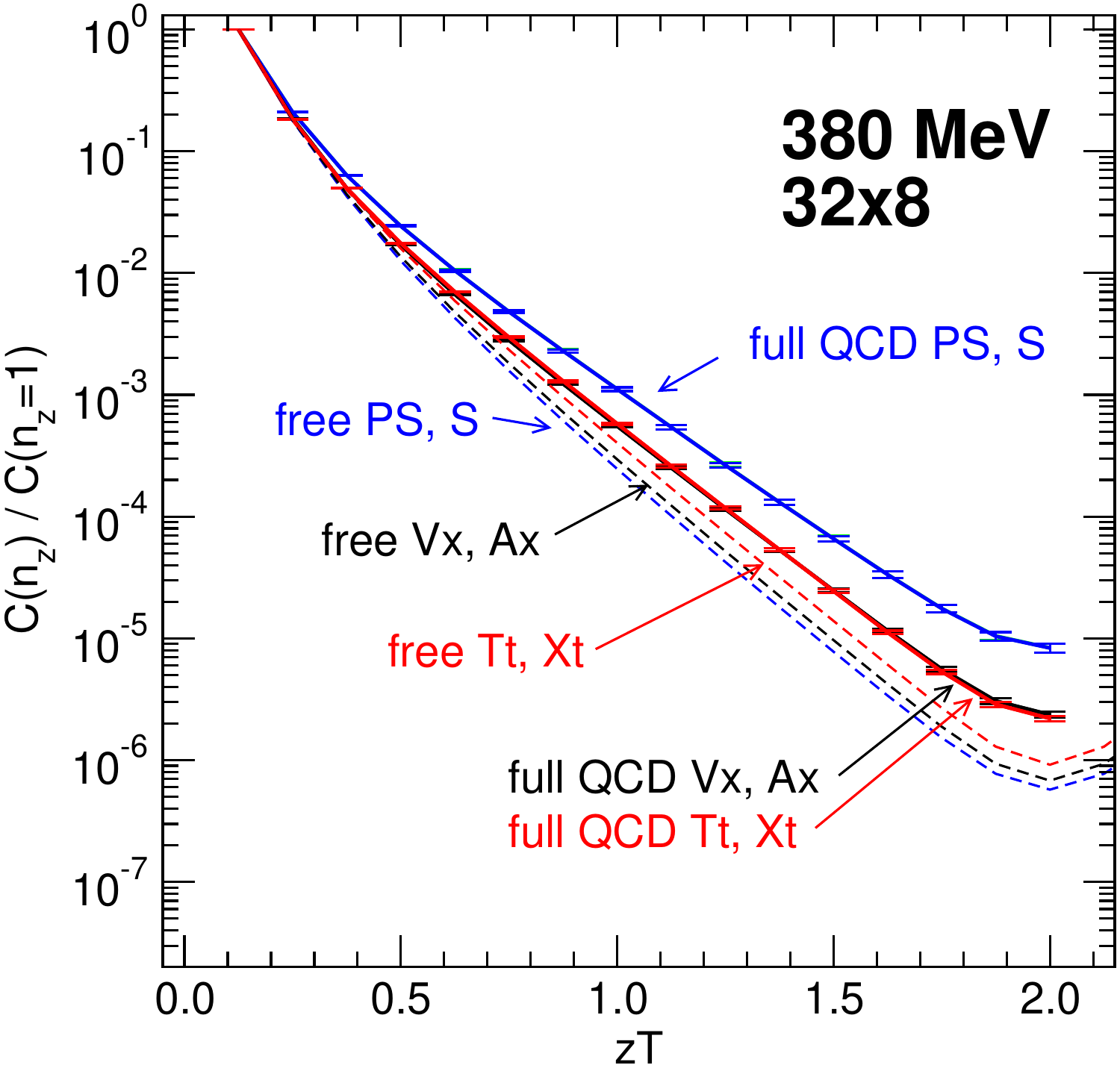} 
  \includegraphics[scale=0.4]{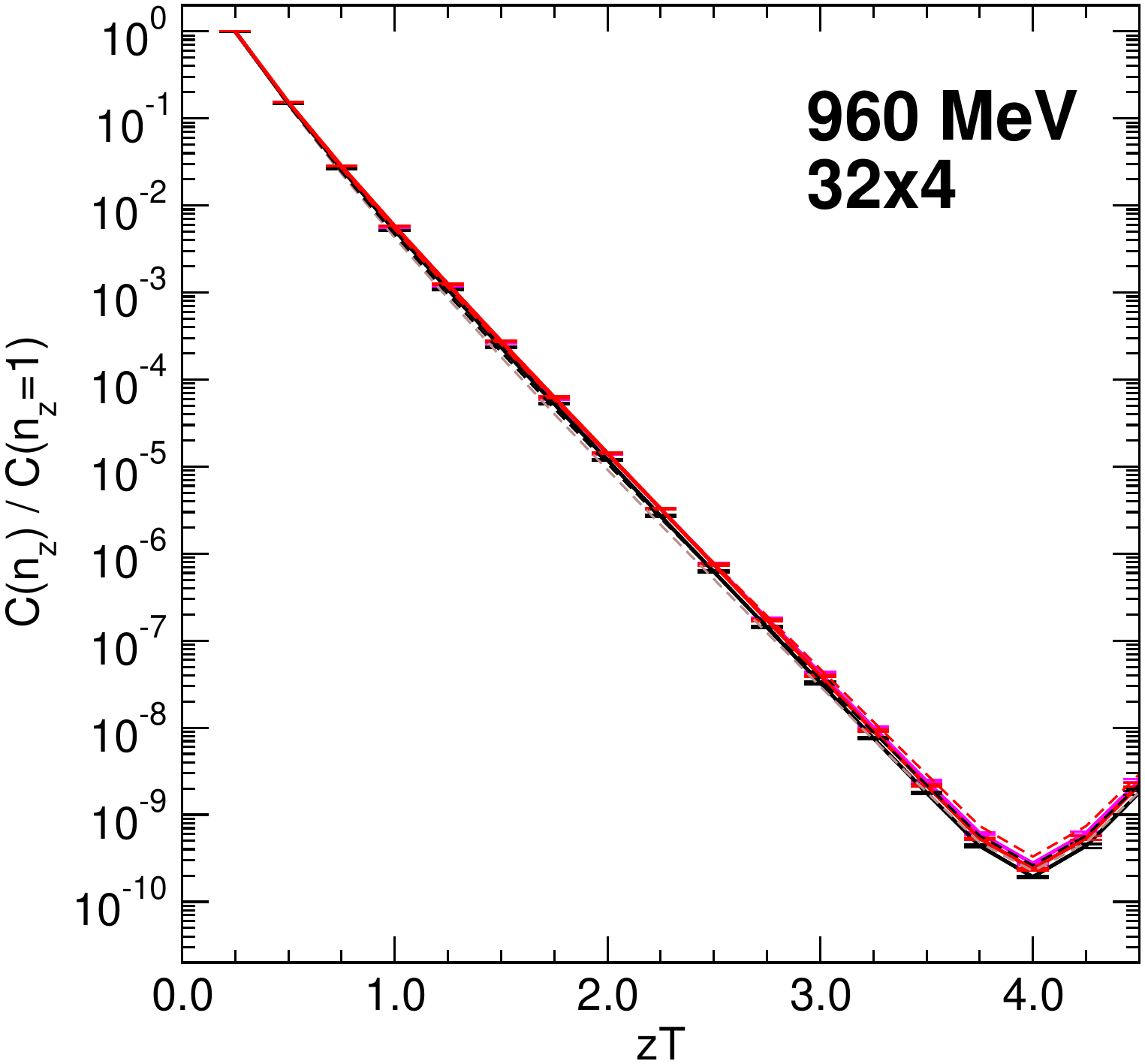}
\caption{
Correlation functions of the bilinears $PS, S, V_x, A_x, T_t, X_t$. The
solid curves represent full QCD calculation and the dashed lines
are correlators calculated with free noninteracting quarks. The Fig. is from
Ref. \cite{R2}.}
\label{fig:e2_withfreedata}
\end{figure}

In Fig. \ref{fig:e2_withfreedata} we show such correlators. Here $PS$ and $S$
denote correlators calculated with the $J=0$ pseudosclar and scalar operators
that are connected by the $U(1)_A$ transformation. Consequently their degeneracy
would indicate restoration of $U(1)_A$ symmetry.
 $V_x, A_x, T_t, X_t$ are
 $J=1$ isovector operators. $V_x$ ($\gamma_1$) and $A_x$ ($\gamma_1 \gamma_5$) are the $x$-components of the vector and axial-vector currents.  
 These two operators are connected by the
 axial part of the $SU(2)_L \times SU(2)_R$ transformation and 
their degeneracy would be a signal of restored
$SU(2)_L \times SU(2)_R$ symmetry. $T_t$ ( $ \gamma_4 \gamma_3$) and
 $X_t$ ($\gamma_4 \gamma_3 \gamma_5)$ operators are connected by the
$U(1)_A$ symmetry and a degeneracy of the corresponding correlators is
required by restored  $U(1)_A$ symmetry. The operators $(A_x, T_t, X_t)$
are connected by the $k=2$ $CS$-transformation (5) which does not mix
these $J=1$ operators with operators of different spin and thus respects
 the rotational $O(3)$ symmetry in Minkowski space at nonzero temperatures.
 A degeneracy of correlators with the $(A_x, T_t, X_t)$ operators would be the evidence of emerged $SU(2)_{CS}$ symmetry. All four operators 
$(V_x, A_x, T_t, X_t)$ are connected by the $SU(4)$ transformation (7)
and a degeneracy of the corresponding correlators would indicate
emergent  $SU(4)$ symmetry. We show the full QCD results (solid lines)
as well as the correlators calculated on the same lattice size with free
noninteracting quarks. Note that the correlators calculated with free
quarks reflect physics at a very high temperature and only chiral
$SU(2)_L \times SU(2)_R$ and $U(1)_A$ symmetries are present in this case.
No $SU(2)_{CS}$ and $SU(4)$ symmetries exist for free quarks.

All correlation functions of chiral partners are degenerate within errors.
In detail, this is the  pair $(V_x,A_x)$ which
reflects the $SU(2)_R \times SU(2)_L$ symmetry.
The $U(1)_A$ symmetry in the scalar $(PS,S)$ pair as well as in the vector channel $(T_t,X_t)$, is manifest. At the lowest available temperature
$220$ MeV ($1.2 T_c$) we observe an obvious multiplet structure: $(PS,S)$ 
 multiplet and the approximately degenerate $(V_x, A_x, T_t, X_t)$ multiplet
reflecting the $SU(2)_{CS}$ and $SU(4)$ approximate symmetries. A residual small
splitting within the latter multiplet is visible. At  larger temperature
$T=380$ MeV ($2.2 T_c$) this residual splitting nearly vanishes.
At these two temperatures we  see the multiplet structures of all
$U(1)_A$, $SU(2)_R \times SU(2)_L$, $SU(2)_{CS}$ and $SU(4)$ symmetries.
At the same time correlators calculated with free quarks demonstrate a
very different multiplet pattern and only chiral symmetries
are manifest. 
 
At the highest temperature of this study, $T \sim 960$ MeV ($5.5 T_c$), the situation has changed significantly:
All full QCD correlators almost  coincide with the corresponding free
correlators (the dashed lines are almost on top of the data points for the full QCD correlators).
Hence at $T \sim 960$ MeV we have reached the region where only chiral
$U(1)_A$ and $SU(2)_L \times SU(2)_R$ symmetries exist and the near coincidence
with the free correlators suggests a gas of quasi-free quarks.

These results  demonstrate that
-- while the chiral symmetries are practically exact -- the $SU(2)_{CS}$ and $SU(4)$ symmetries are not exact. 
Now we introduce a measure for the symmetry  breaking and find  the temperature range where the
symmetry is appropriate.

In general a symmetry is established via its multiplet structure.
For any multiplet structure a crucial parameter is the ratio of the
splitting within a multiplet to the distance between multiplets.
Consequently, in our case the 
breaking of $SU(2)_{CS}$ and $SU(4)$ can be identified through the
parameter

\begin{equation}\label{def_kappa}
  \kappa = \frac{|C_{A_x} - C_{T_t}|}{|C_{A_x} - C_{S}|}.
\end{equation} 
If $\kappa \ll 1$, then we can declare
 an approximate or -- if zero -- an exact
symmetry.
If $\kappa \sim 1$, the symmetry is absent.
From the free quark analytical calculations one finds  $\kappa \sim 1$ 
\cite{R2}, which stresses  that there is no chiral-spin
symmetry for free quarks. 

\begin{figure}
  \centering
  \includegraphics[scale=0.5]{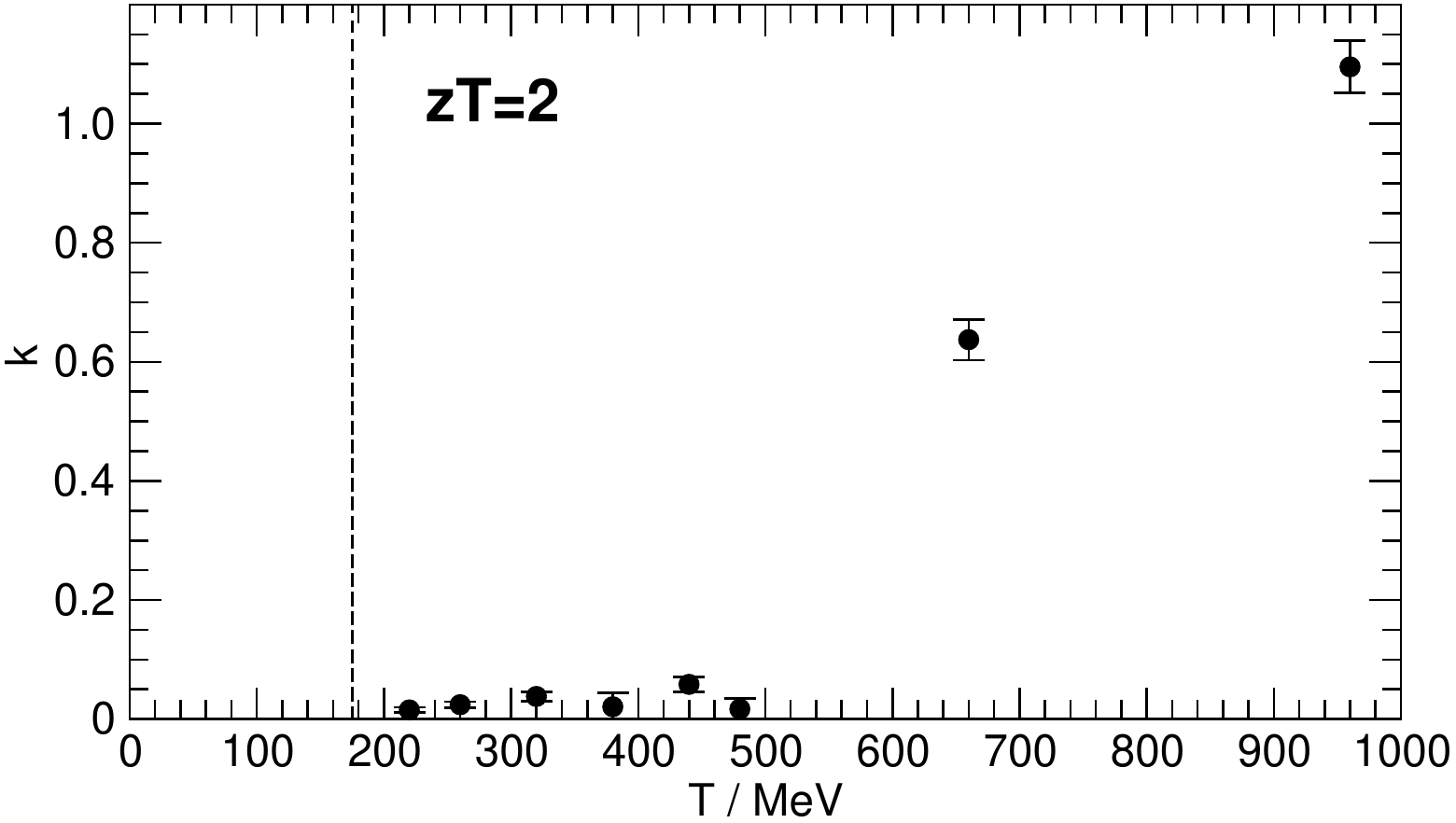}
  \caption{The symmetry breaking parameter $\kappa$ defined in (\ref{def_kappa}) evaluated at $zT=2$ for different temperatures.
           The dashed line represents $T_c$.}
  \label{fig:kappa_zT2}
\end{figure}

In Fig. \ref{fig:kappa_zT2} we show the evolution of the symmetry breaking parameter $\kappa$ as
a function of temperature at $zT=2$.
The value of $\kappa$ is less than 5 \% for all ensembles with $T \sim 220$ -- $500$ MeV.
This implies that the symmetries that we observe in the range between
$T \sim 220$ MeV and   $500$ MeV are well  pronounced. These symmetries
of spatial correlators reflect symmetries of the QCD action since correlators
are driven only by the action of the theory.

At temperatures between $T \sim 500$ MeV and  $T \sim 660$ MeV we notice a drastic increase
of the symmetry breaking parameter $\kappa$ to values of the order $\sim 1$.
We conclude that QCD exhibits approximate $SU(2)_{CS}$  and $SU(4)$
symmetries in the
temperature range between $T \sim 220$ -- $500$ MeV with symmetry breaking
less than 5\% as measured with $\kappa$.
This suggests that the $SU(2)_{CS}$ symmetric regime begins just after
the $SU(2)_R \times SU(2)_L$ restoration crossover.

\begin{figure}
  \centering
  \includegraphics[scale=0.45]{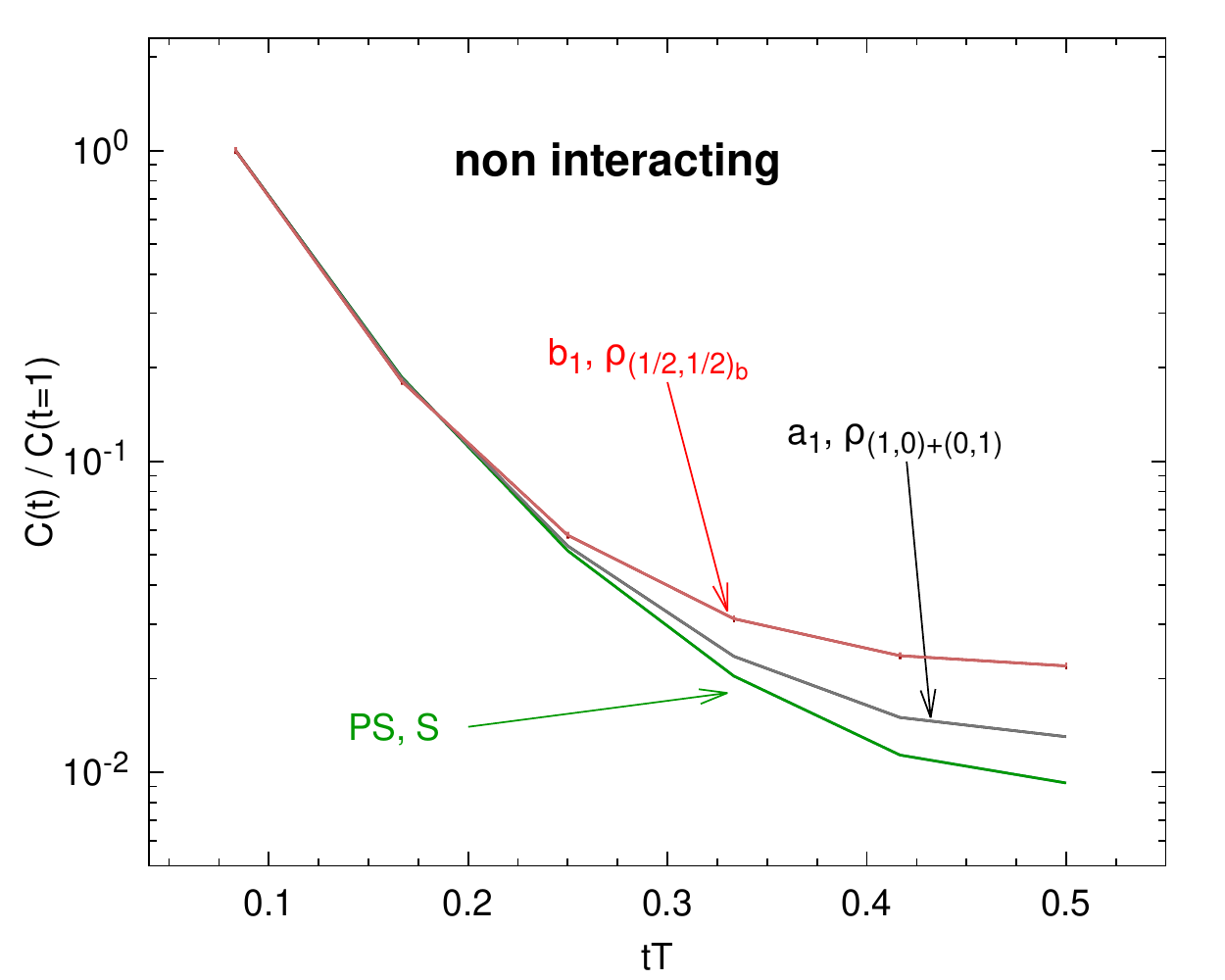} 
  \includegraphics[scale=0.45]{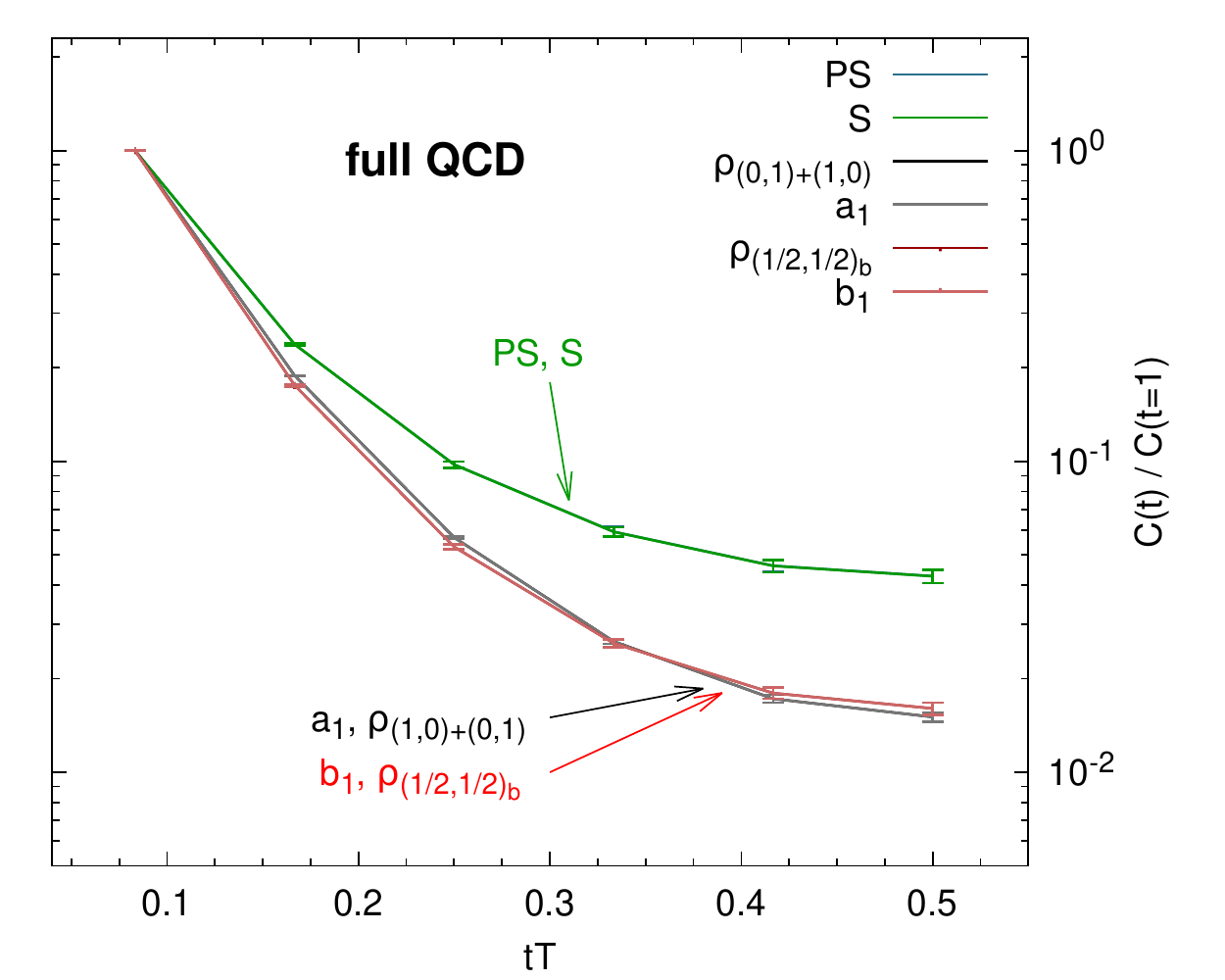} 
\caption{ Temporal correlation functions for $12 \times 48^3$
lattices. The l.h.s. shows correlators calculated with free
noninteracting quarks with manifest $U(1)_A$  and $SU(2)_L \times SU(2)_R$
symmetries. The r.h.s. presents full QCD results at a temperature $1.2 T_c$,
which shows multiplets of all  $U(1)_A$, $SU(2)_L \times SU(2)_R$, $SU(2)_{CS}$  and $SU(4)$ groups. The Fig. is from Ref. \cite{R3}.
}
\label{tcorr}
\end{figure}

On the right side of Fig.~\ref{tcorr} we show temporal correlators

\begin{equation}
C_\Gamma(t) = \sum\limits_{x, y, z}
\braket{\mathcal{O}_\Gamma(x,y,z,t)
\mathcal{O}_\Gamma(\mathbf{0},0)^\dagger},
\label{eq:momentumprojection}
\end{equation}
at a temperature $T = 1.2 T_c$ \cite{R3}
where $\mathcal{O}_\Gamma(x,y,z,t)$ is an operator that creates a
quark-antiquark pair  with fixed quantum numbers. Summation over
$x,y,z$ projects out the hadron rest frame. 

Transformation properties of the local  $J=1$  quark-antiquark bilinears 
$\mathcal{O}_\Gamma(x,y,z,t)$
 with respect to $U(1)_A$, $SU(2)_L \times SU(2)_R$, $SU(2)_{CS}, k=4$  and  $SU(4)$  are given in  
Fig.~\ref{F3}.
Emergence of the respective symmetries is signalled by the degeneracy of
the correlators (\ref{eq:momentumprojection}) calculated with operators
that are connected by the corresponding transformations.

On the l.h.s of Fig.~\ref{tcorr} we demonstrate correlators
calculated with noninteracting quarks on the same lattice.
They represent the QGP at a very high temperature where due to asymptotic freedom
the quark-gluon interaction can be neglected.
Dynamics of free quarks are governed by the Dirac
equation and only $U(1)_A$ and $SU(2)_L \times SU(2)_R$ chiral symmetries
exist. A qualitative difference between the pattern on the l.h.s.   and the pattern on the r.h.s of Fig.~\ref{tcorr}
is obvious. In the latter case we clearly see 
$SU(2)_{CS}$ and $SU(4)$ symmetries. The temporal correlators
are directly connected to  measurable spectral density.
$SU(2)_{CS}$ and $SU(4)$ symmetries of the t-correlators imply the same 
symmetries of spectral densities.

The emergent $SU(2)_{CS}$ and $SU(4)$ symmetries seen at $T_c - 3 T_c$
suggest that the physical degrees of freedom at these temperatures
are chirally symmetric quarks bound into color singlets by the
chromoelectric field. The magnetic effects are either absent or strongly
suppressed. Upon increasing temperature above $T_c$ a strength of the confining electric field smoothly diminishes and at $T > 3 T_c$ it becomes
sufficiently small so that the emergent $SU(2)_{CS}$ and $SU(4)$ symmetries
disappear.

\section{Conclusions}

In the range $T_c - 3 T_c$   we observe  formation of multiplets
in  correlators that indicate  emergent approximate symmetries described by the chiral-spin 
$SU(2)_{CS}$ and $SU(4)$ groups.   
These symmetries include the chiral $U(1)_A$ and $SU(2)_L \times SU(2)_R$  as subgroups.  These are not symmetries of the free Dirac action. In a given reference frame, which in our case is the medium rest frame,
the  chromoelectric interaction is invariant under both $SU(2)_{CS}$ and $SU(4)$ transformations,
while the chromomagnetic interaction as well as the quark kinetic term break them. 

The emergence of these symmetries in the $T_c$ - $3 T_c$  window
suggests that the chromomagnetic field disappears or is strongly
suppressed, while the confining chromoelectric
field is still active.
This implies that the
physical degrees of freedom are chirally symmetric quarks bound by the
chromoelectric interaction into color-singlet objects without chromomagnetic effects.
While we do not advocate any microscopic description of these ultrarelativistic objects,
they are reminiscent of ``strings".
We refer the $SU(2)_{CS}$ and $SU(4)$ symmetric regime at temperatures
$T_c$ -- $3 T_c$ as the ``Stringy Fluid" to  emphasize the possible  nature of the
objects - chirally symmetric quarks bound by the  electric field \cite{G4,R2}.

\begin{figure}
\centering
\includegraphics[scale=0.3]{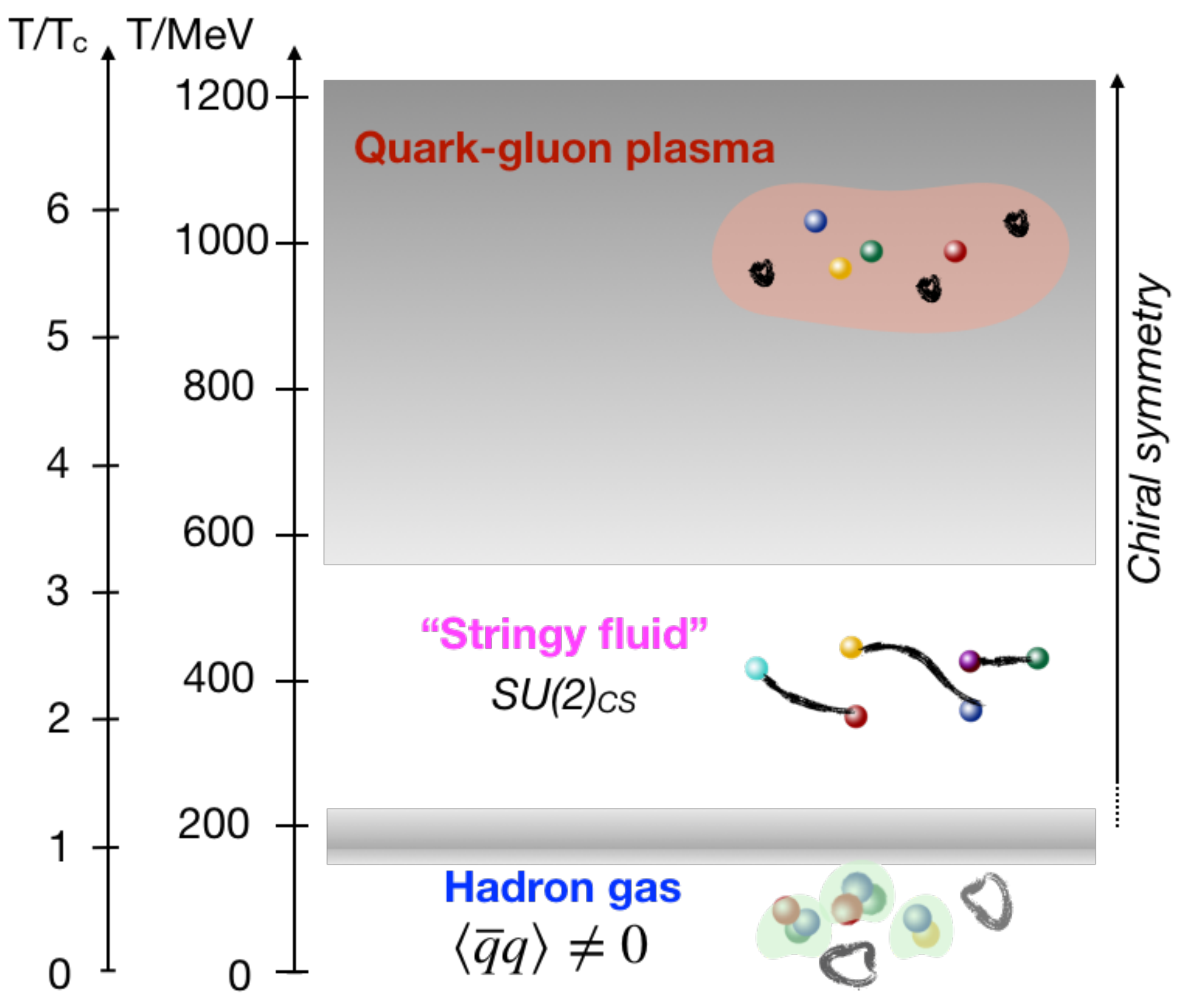} 
\caption{
Sketch for the temperature evolution of the QCD effective degrees of freedom.
The Fig. is from Ref. \cite{R2}.}
\label{fig:sketch}
\end{figure}

These results suggest the following three
regimes of QCD.  At  temperatures below
the pseudocritical temperature $T_c$ QCD matter is a Hadron Gas where all chiral
symmetries are spontaneously broken by the non-zero quark condensate. 
From the Hadron Gas regime   
there is a   crossover to a  regime  
with approximate $SU(2)_{CS}$ chiral-spin symmetry, where quarks are predominantly bound by
the chromoelectric field. This crossover either coincides or is close
to the chiral $SU(2)_L \times SU(2)_R$ restoration crossover.
In the range above  $ \sim 3 T_c$  there is a fast increase
of chiral-spin symmetry breaking:
the confining electric interaction becomes small relative to the quark kinetic term.
Finally, up to $ \sim 5 T_c$ GeV there is an evolution to a weakly interacting QGP, where the relevant 
symmetries are the full set of chiral symmetries and quarks behave as if
they were free. Fig.~\ref{fig:sketch} provides an illustrative sketch of this temperature
evolution for the effective degrees of freedom of QCD.
All three regimes of QCD, the Hadron Gas, the Stringy Fluid and the QGP differ
by symmetries, degrees of freedom and physical properties. E.g., while
the viscosity of the Hadron Gas and of QGP is very large, as in an ideal gas, the Stringy Fluid, as it follows from experiments,
is a perfect one. One can speculate that this feature is related to
a relatively large size of "strings" in the Stringy Fluid.

Our conclusions about three regimes of QCD  are based only on symmetry classification
of the QCD Lagrangian and lattice results obtained at zero
chemical potential. At a nonzero chemical potential the
lattice Monte-Carlo is not efficient.
The quark chemical potential term $ \mu \psi (x)^\dagger \psi(x)$ 
in the  QCD action

\begin{equation}
S = \int_{0}^{\beta} d\tau \int d^3x
\overline{\psi}  [ \gamma_{\mu} D_{\mu} + \mu \gamma_4 + m] \psi,
\end{equation}
\noindent
 is $SU(2)_{CS}$  and $SU(2N_F)$
invariant \cite{G4}. Consequently $SU(2)_{CS}$  and $SU(2N_F)$ symmetries
observed on the lattice at $ \mu =0$ should also persist in a medium with
a finite chemical potential.

We will finish this presentation with a historical analogy.
Christopher Columbus had a theory that moving to the west through
the Ocean he would reach India. He was able to convince Ferdinand
and Isabella of Castile to finance his "experimental research". And indeed
after three months of  adventures he reached a land that was perfectly
seen to him as India. But actually he discovered America. Inspite of the
erroneous theoretical prediction and interpretation of results his discovery
was of great importance.
end{equation}

\end{document}